\documentclass[ twocolumn, twocolappendix]{aastex631}

\DeclareUnicodeCharacter{02BC}{'}

\shortauthors{Search for Fast Radio Bursts and radio pulsars from PULXs}

\begin{document}

\title{Search for Fast Radio Bursts and radio pulsars from pulsing Ultraluminous X-ray Sources}

\correspondingauthor{Na Wang, Rui Luo, Shuangqiang Wang}
\email{na.wang@xao.ac.cn, rui.luo@gzhu.edu.cn, wangshangqiang@xao.ac.cn}

\nocollaboration{1}

\author[0000-0002-1052-1120]{Juntao Bai}
\affiliation{Xinjiang Astronomical Observatory, Chinese Academy of Sciences, Urumqi, Xinjiang 830011, China}
\affiliation{School of Astronomy and Space Science, University of Chinese Academy of Sciences, Beijing 100049, China}

\author[0000-0002-9786-8548]{Na Wang}
\affiliation{Xinjiang Astronomical Observatory, Chinese Academy of Sciences, Urumqi, Xinjiang 830011, China}
\affiliation{Key Laboratory of Radio Astronomy, Chinese Academy of Sciences, Urumqi, Xinjiang 830011, China}

\author[0000-0002-4300-121X]{Rui Luo}
\affiliation{Department of Astronomy, School of Physics and Materials Science, Guangzhou University, Guangzhou, 510006, China}

\author[0000-0001-9036-8543]{Wei-Yang Wang}
\affiliation{School of Astronomy and Space Science, University of Chinese Academy of Sciences, Beijing 100049, China}

\author[0000-0002-9618-2499]{Shi Dai}
\affiliation{CSIRO Space and Astronomy, PO Box 76, Epping, NSW 1710, Australia}
\affiliation{Western Sydney University, Locked Bag 1797, Penrith South DC, NSW 2751, Australia}

\author{Songbo Zhang}
\affiliation{Purple Mountain Observatory, Chinese Academy of Sciences, Nanjing 210023, China}

\author{Shiqian Zhao}
\affiliation{Department of Astronomy, School of Physics and Materials Science, Guangzhou University, Guangzhou, 510006, China}

\author[0000-0003-4498-6070]{Shuangqiang Wang}
\affiliation{Xinjiang Astronomical Observatory, Chinese Academy of Sciences, Urumqi, Xinjiang 830011, China}
\affiliation{CSIRO Space and Astronomy, PO Box 76, Epping, NSW 1710, Australia}
\affiliation{Key Laboratory of Radio Astronomy, Chinese Academy of Sciences, Urumqi, Xinjiang 830011, China}

\nocollaboration{8}

\begin{abstract}

We conducted targeted fast radio burst (FRB) and pulsar searches on eight pulsing ultraluminous X-ray sources (PULXs) using the Five-hundred-meter Aperture Spherical Radio Telescope (FAST) and the Parkes 64-meter Radio Telescope (Murriyang) to investigate whether PULXs could be progenitors of FRBs.
FAST carried out 12 observations of four PULXs, totaling 8 hours, while Parkes conducted 12 observations of the remaining four PULXs, totaling 11 hours.
No significant signals were detected through single-pulse and periodic searches, covering a dispersion measure (DM) range of 0–5000 pc cm$^{-3}$, placing stringent upper limits on the radio flux density from these sources. 
The results imply that accretion processes and dense stellar winds in PULXs likely suppress or attenuate potential coherent emission in radio band. Additionally, the beaming factor and luminosity of FRBs associated with PULXs, as well as the highly relativistic and magnetized nature of their outflows, may limit detectability. Non-detection yielded from the observations covering the full orbital phases of PULXs can also constrain the theoretical models that link FRB emission to highly magnetized neutron stars in binary systems.

\end{abstract}


\section{INTRODUCTION}

In the dynamic universe, some extremely fast transients can be captured under high-time resolution digitised system, namely Fast Radio Bursts (FRBs), which was firstly discovered in 2007 \citep{Lorimer2007Sci}. This kind of radio flashes are typically characterized as single pulses with durations from dozens of microseconds to tens of milliseconds and remarkable dispersion exceeding the line-of-sight contributions by the Milky Way \citep{CHIME/FRB2021ApJS, Xu2023Univ}. At present, FRBs have been merely detected at radio bands, particularly in the range of 120 MHz \citep{Pastor-Marazuela+21Nat} to 8 GHz \citep{Gajjar+18ApJ}, and no detections have been made at other wavelengths.
Despite extensive observational efforts, no high-energy astrophysical events have so far been observed in association with FRBs \citep{Men2019MNRAS, Madison2019ApJ, Hilmarsson2020MNRAS, Pilia2020ApJ, Scholz2020ApJ, Laha2022ApJ, Cook2024ApJ, Pearlman2025NatAs}.
Nevertheless, a small fraction of FRB sources have been observed to be repeatable, raising the open question of whether repeating and one-off bursts share distinct origins \citep{Chime/Frb2023ApJ}.

The origin(s) of FRBs remains enigmatic despite numerous theoretical models. Based on the short duration ($\sim$ ms) and extremely high inferred brightness temperature, coherent emission processes are required to produce such bursts. Magnetars have become the most plausible progenitors owing to a emerging of observational breakthroughs \citep{Zhang2020Natur}, in particular, the discovery of the Galactic FRB-like event. In April 2020, an exceptionally bright radio burst (FRB 200428) with fluences of $\sim$ million Jy\,ms was detected from the Galactic magnetar SGR 1935+2154 by CHIME and STARE2 simultaneously \citep{CHIME/FRB2020Natur.587, Bochenek2020Natur}. This discovery demonstrates that magnetars are capable of producing FRBs both within our own galaxy and at extragalactic distances.
Several models have been proposed by deriving coherent radiation mechanism in magnetar \citep{2017MNRAS.468.2726K,2018ApJ...868...31Y,2019MNRAS.485.4091M,2019ApJ...876L..15W,2020MNRAS.498.1397L,2022ApJ...927....2K,2022ApJ...927..105W,2023ApJ...943...47L,2023MNRAS.522.2448Q,2024ApJ...972..124Q}.
While these theories provide valuable insights, further observational and theoretical studies are required to confirm and refine these models.

Interestingly, recent studies proposed that some observed FRBs might arise from relativistic magnetized shocks or magnetic reconnection events in highly energetic systems \citep{Zhang2023RvMP}. Such mechanisms are not exclusive to magnetar engines and could be explained by short-lived relativistic outflows from accreting black hole or neutron star binaries \citep{Li+23Nat}. Among these, ultra-luminous X-ray (ULX) sources, powered by super-Eddington accretion onto stellar-mass compact objects, present a compelling possibility \citep{Sridhar2021ApJ}. The spatial distribution of ULXs within host galaxies shows striking similarities to FRB locations, and their super-Eddington accretion processes can naturally account for the extreme luminosities observed in FRBs. This connection highlights the potential role of ULXs in bridging our understanding of FRB progenitors and their energy sources.

Some ULXs exhibit periodic X-ray pulsations ranging from seconds to minutes \citep{King2020MNRAS}, so that the sources are named as pulsing ULXs \citep[PULXs; e.g.,][]{Bachetti2014Natur}.
It is initially suggested that PULXs may have strong magnetic field (like the magnetar), which can evade the Eddington limit by reducing the electron-scattering cross-section \citep{Eksi2015MNRAS, Mushtukov2015MNRAS}. PULX might be a magnetar in an accreting system \citep{Tong2019MNRAS}, which also favored the idea that magnetars were descendants of high mass X-ray binaries \citep{Bisnovatyi-Kogan2015ARep}. 

Driven by the hypothesis that ULXs could be potential progenitors of FRBs and might share a close association \citep{Sridhar2021ApJ}, we carried out a targeted FRB search for known PULXs using the Five-hundred-meter Aperture Spherical Radio Telescope (FAST) and the Parkes 64-meter Radio Telescope (also known as Murriyang). With extremely high sensitivity of FAST \citep{nan2011IJMPD, li2018IMMag} and the ultra-wide-bandwidth of Parkes telescope \citep{Hobbs2020PASA}, our study will provide more information on the understanding of the origins of FRBs. PULXs may host radio pulsars, which has not yet been detected at radio waveband \citep{Kaaret2017ARA&A}. Therefore, we also search for radio pulsars in PULXs.
In this paper, we present details on the observations in Section \ref{sec2}, the results are presented in Section \ref{sec3}, and we discuss and summarize the results in Section \ref{sec4}.

\begin{table*}[!htp]
    \caption{ Observed properties of 8 PULXs. }
    \label{table1}
    \begin{tabular}{lcccccc}
    \hline\hline
    Name                    & Spin period 	& $\dot{\nu}$          	  & P$_{orb}$  &  L$_{max}$                         &  Distance   &   References  \\
                            &  (s)	        & ($10^{-10}$ s$^{-2}$)   &  (day)     &  (10$^{39}$ erg s$^{-1}$)          &   (Mpc)     &               \\
    \hline	
    M51 ULX-7                   &   2.8         &   2.8          &   $\sim$2     &    7          &   8.6        &  (1), (2)   \\
    M51 ULX-8 $^{\dagger}$      &   --          &   --           &   8-400      &    2          &   8.6        &  (3)  \\
    NGC5907 ULX-1               &   1.13        &   38           &   5.3        &    $\sim$100  &   17.1       &  (4)--(6)  \\
    Swift J0243.6+6124          &   9.86        &   2.2          &   28.3        &    $\gtrsim$1.5     &  0.005 &  (7)  \\
    NGC1313 PULX                &   $\sim$765.6 &  --            &   --          &    1.6        &   13.85      &  (8)  \\
    NGC300 ULX-1                &   $\sim$31.6  & 5.6            &  \textgreater 290  &   4.7  &   1.88       &  (9), (10)  \\
    NGC7793 P13                 &   0.42        & 2              &  63.9         &    5          &   3.6        &  (11)--(13)  \\
    SMC X-3                     &  7.7   &  0.69    & $\sim$45        &    2.5        &   0.062      &  (14)--(17)  \\
    \hline
    \end{tabular}
    \tablecomments{$^\dagger$ Cyclotron resonant scattering feature (CRSF) detected, but no pulsations. \\
    References: (1) \citealt{Rodr2020ApJ}; (2) \citealt{Vasilopoulos2020MNRAS}; (3) \citealt{Brightman2018NatAs}; (4) \citealt{Walton2016ApJ}; (5) \citealt{Israel2017Sci}; (6) \citealt{furst2023A&A}; (7) \citealt{Doroshenko2018A&A}; (8) \citealt{Trudolyubov2008MNRAS}; (9) \citealt{Carpano2018MNRAS}; (10) \citealt{Heida2019ApJ}; (11) \citealt{Furst2016ApJ}; (12) \citealt{Israel2017MNRAS}; (13) \citealt{furst2021A&A}; (14) \citealt{Edge2004MNRAS}; (15) \citealt{Corbet2004AIPC}; (16) \citealt{Tsygankov2017A&A}; (17) \citealt{Townsend2017MNRAS}. }
\end{table*}

\section{OBSERVATIONS} \label{sec2} 

About 12 PULXs have been discovered to date \citep{King2023NewAR}. Table \ref{table1} summarizes the parameters of the 8 PULXs we selected for this study, considering the sky coverage of the FAST and Parkes telescopes.  We conducted observations of these 8 PULXs using the FAST and Parkes telescopes individually, and the observational details are presented in Table \ref{table2}.  
For the FAST observations, we utilized the central beam of the 19-beam receiver, covering a frequency range of 1.05 to 1.45 GHz \citep{jiang2020RAA}, to observe four of the eight sources. Each PULX was observed three times, resulting in a total of 12 observation sessions, with each session lasting 2400 seconds. The data were recorded in pulsar search mode with a bandwidth of 500 MHz over 4096 channels, a sample interval of 49.152 $\mu$s, and 8-bit digitization.  For the Parkes observations, we used the ultra-wide-bandwidth, low-frequency receiver (UWL, \citealt{Hobbs2020PASA}) covering the frequency range of 704 to 4032 MHz to observe the remaining four PULXs. Each PULX was also observed three times, with session durations ranging from 2671 to 4821 seconds, resulting in 12 observation sessions in total. The data were recorded in pulsar search mode with a bandwidth of 3328 MHz over 6656 channels, a sample interval of 64 $\mu$s, and 8-bit format.

\section{Data Analysis and Results}\label{sec3}

We utilized the \texttt{PRESTO} analysis suite to search for both single pulses and periodic radio signals in each observation, following the methodology outlined in \citet{ransom2002AJ}. In the initial processing stage, narrowband and transient radio frequency interference (RFI) were identified and mitigated using the \texttt{RFIFIND} utility from \texttt{PRESTO}.
Before commencing the search process, data were de-dispersed across a broad range of Dispersion Measures (DM). De-dispersion was performed within the range of 0 to 5000 pc cm$^{-3}$ for all eight PULXs. 
For the periodic pulse signal search, we used the entire frequency band for both Parkes and FAST observations, as pulsar radio emissions are broadband. However, since pulsars typically weaken at high frequencies or may exhibit significant scattering at low frequencies (e.g., \citealt{2024ApJ...967L..16L}), we further divided the UWL observation into four sub-bands of 832 MHz each for the periodic pulse signal search. For the single pulse search, considering the narrowband nature of FRBs, we divided the UWL frequency band into the following sub-bands: 1 $\times$ 3328 MHz, 2 $\times$ 1664 MHz, 4 $\times$ 832 MHz, 8 $\times$ 416 MHz, 13 $\times$ 256 MHz, and 26 $\times$ 128 MHz (e.g., \citealt{Kumar2022MNRAS}), while the full frequency band of FAST observations was used (e.g., \citealt{2022Natur.606..873N}).
This segmentation allowed for a more focused search across different frequency ranges. To enhance dedispersion accuracy while maintaining computational efficiency, a fine DM spacing was used for low DM values and a slightly coarser step for higher DM values (d(DM) = 0.05-3 pc cm$^{-3}$), following the guidance from \texttt{DDplan.py} \citep{ransom2011ascl}.

\begin{table*}
    \caption{Observational parameters for the 8 PULXs.}
    \label{table2}
    \hspace*{-1.2cm}
    \begin{tabular}{lccccccc}
    \hline\hline
    Name        &   RA 	            &   DEC	      &   MJD &   Date          & Length   & Telescope  &  Receiver 	\\ 
                &  (J2000)   	    &  (J2000)    &       &  (yyyy-mm-ss)   & (s)      &            &     	        \\
    \hline	
    M51 ULX-7           &   13:30:01.02     &   +47:13:43.80     &   59487.31   &  2021-09-30  &    2400    & FAST      &  19-beam    \\
                        &                   &                    &   59488.30   &  2021-10-01  &    2400    & FAST      &  19-beam    \\
                        &                   &                    &   59489.26   &  2021-10-02  &    2400    & FAST      &  19-beam    \\
    M51 ULX-8           &   13:30:07.55     &   +47:11:06.10     &   59489.18   &  2021-10-02  &    2400    & FAST      &  19-beam    \\
                        &                   &                    &   59496.18   &  2021-10-09  &    2400    & FAST      &  19-beam    \\
                        &                   &                    &   59503.25   &  2021-10-16  &    2400    & FAST      &  19-beam    \\
    NGC5907 ULX-1       &   15:15:58.62     &   +56:18:10.30     &   59487.34   &  2021-09-30  &    2400    & FAST      &  19-beam    \\
                        &                   &                    &   59489.30   &  2021-10-02  &    2400    & FAST      &  19-beam    \\
                        &                   &                    &   59491.31   &  2021-10-04  &    2400    & FAST      &  19-beam    \\
    Swift J0243.6+6124  &   02:43:40.44     &   +61:26:03.73     &   59486.75   &  2021-09-29  &    2400    & FAST      &  19-beam    \\
                        &                   &                    &   59496.76   &  2021-10-09  &    2400    & FAST      &  19-beam    \\
                        &                   &                    &   59506.71   &  2021-10-19  &    2400    & FAST      &  19-beam    \\
    NGC1313 PULX        &   03:17:47.59     &   -66:30:10.20     &   59491.82   &  2021-10-04  &    4821    & Parkes    &  UWL     \\
                        &                   &                    &   59565.38   &  2021-12-17  &    3214    & Parkes    &  UWL     \\
                        &                   &                    &   59642.34   &  2022-03-04  &    3302    & Parkes    &  UWL     \\
    NGC300 ULX-1        &   00:55:04.86     &   -37:41:43.70     &   59488.48   &  2021-10-01  &    3287    & Parkes    &  UWL     \\
                        &                   &                    &   59564.30   &  2021-12-16  &    3301    & Parkes    &  UWL     \\
                        &                   &                    &   59657.17   &  2022-03-19  &    3301    & Parkes    &  UWL     \\
    NGC7793 P13         &   23:57:50.90     &   -32:37:26.60     &   59488.44   &  2021-10-01  &    3302    & Parkes    &  UWL     \\
                        &                   &                    &   59564.38   &  2021-12-16  &    2864    & Parkes    &  UWL     \\
                        &                   &                    &   59657.13   &  2022-03-19  &    3301    & Parkes    &  UWL     \\
    SMC X-3             &   00:52:05.70     &   -72:26:04.00     &   59491.78   &  2021-10-04  &    3302    & Parkes    &  UWL     \\
                        &                   &                    &   59564.34   &  2021-12-16  &    3301    & Parkes    &  UWL     \\
                        &                   &                    &   59657.30   &  2022-03-19  &    2671    & Parkes    &  UWL     \\
    \hline
    \end{tabular}
\end{table*}

\subsection{Single Pulse Search}\label{sec3.1}

To identify dispersed single pulses and bursts, we processed the dedispersed time series following the approach detailed in \citet{cordes2003ApJ}. The analysis was conducted using \texttt{single\_pulse\_search.py} from the \texttt{PRESTO} software package. The DM range and step sizes were consistent with those previously described. The native time resolution of 0.05 ms was retained throughout the analysis to preserve sensitivity to short-duration pulses. Candidate pulses were identified by applying a signal-to-noise ratio (S/N) threshold of 7 (e.g., \citealt{Zhi2024ApJ}), chosen to effectively differentiate astrophysical signals from background RFI.  

To extend sensitivity to broader pulses, we employed matched filtering with boxcar functions spanning widths from 1 to 30 samples. For longer-duration bursts, the dedispersed time series were downsampled by aggregating adjacent samples into groups of 2, 4, 8, 16, 32, and 64. Each set of downsampled time series was then searched using the same set of boxcar filters, providing sensitivity to signals with durations up to the maximum boxcar length of 30 samples multiplied by the maximum downsampled sample time.

For each PULXs, we consolidated search candidates by clustering them within a 30-second time window and a DM range of 500 pc cm\(^{-3}\). The resulting two-dimensional (2D) DM-time plots were manually reviewed to identify significant clusters. Genuine astrophysical pulses were expected to form distinct clusters at specific DM values and exhibit S/N values exceeding the threshold.   After cross-checking the 2D plots covering the relevant DMs and the dynamic spectra of significant single pulse candidates, we conclude that no astrophysical single pulses were found at an S/N threshold of 7.

\subsection{Periodic Pulse Search}\label{sec3.2}

To search for periodic pulse signals, we employed a systematic approach using the \texttt{PRESTO} software package \citep{ransom2011ascl}. First, a Fast Fourier Transform (FFT) was applied to the dedispersed time-series data with \texttt{realfft}, converting the signal from the time domain to the frequency domain. Next, acceleration searches were performed using \texttt{accelsearch}, with the \textit{zmax} parameter set to 200. This setting ensures that signals exhibiting significant frequency drifts caused by orbital acceleration remain coherent and detectable in the power spectrum.
We conducted a blind search over a wide range of trial spin periods, rather than limiting it to a restricted range, due to the large uncertainties in the spin periods of some PULXs. Note that the detectable periods are primarily constrained by the data sampling rate and observation duration.
Candidate signals surpassing the S/N threshold were folded to evaluate potential periodic pulsations from both the targeted PULXs and any unknown sources within the data. The folding analysis, conducted with \texttt{prepfold}, produced pulse profiles for detailed inspection. Particular attention was given to candidates near the expected spin periods of the PULXs, extrapolated to the observation epoch using precise X-ray timing results (see Table \ref{table1}). Despite this comprehensive analysis, no periodic radio signals of astrophysical origin were identified at the established S/N threshold of 7.

In our search for radio pulses from eight PULXs, we also applied the Fast Folding Algorithm (FFA), a phase-coherent search method optimized for detecting long-period, periodic signals. FFA was chosen because PULXs often exhibit relatively long spin periods, typically exceeding 0.4 second. Traditional search methods such as FFT become less sensitive to such long-period signals, particularly in the presence of low-frequency noise, making FFA a more suitable approach \citep{Morello2020MNRAS}. The FFA works by folding the time-series data over a wide range of trial periods, generating integrated pulse profiles that are examined for periodic signals. Unlike FFT, FFA retains phase information throughout the folding process, making it more sensitive to weak, long-period signals that may otherwise be missed. 
To ensure a comprehensive and sensitive search, we further applied FFA following the guidelines from the \texttt{riptide} FFA documentation\footnote{https://riptide-ffa.readthedocs.io/en/latest/index.html}. Our search was conducted across an extensive grid of trial periods, covering a broad range from 0.1 s to 50 s with fine period resolution to account for potential drifts. For NGC 1313 PULX, we extended the search range further, from 0.1 to 800 s, to accommodate the possibility of even longer spin periods. Multiple diagnostic plots were examined to identify low-duty-cycle signals that might be missed in a standard FFT search.
Despite the thorough application of FFA, no significant periodic radio signals were detected in our observations.

\subsection{Upper limits of the flux density and fluence}\label{sec3.3}

We can determine upper limits on the flux density for the eight PULXs based on the results from both periodic radio pulsation and single pulse searches. The minimum detectable radio flux density, \( S_{\text{min}} \), is estimated using the radiometer equation \citep{lorimer2006MNRAS}:
\begin{equation}
  S_{\text{min}} = (S/N)_{\text{min}} \frac{\beta T_{\text{sys}}}{G \sqrt{n_p t_{\text{obs}} \Delta f}} \sqrt{\frac{W}{P-W}}
  \label{eq1}
\end{equation}
Here, \((S/N)_{\text{min}}\) represents the signal-to-noise threshold required for pulsation detection, \(\beta\) is a dimensionless factor accounting for quantization losses, \(T_{\text{sys}}\) is the system noise temperature (K), \(G\) is the telescope gain (K Jy\(^{-1}\)), \(n_p\) denotes the number of summed polarizations, \(t_{\text{obs}}\) is the observation duration (s), \(\Delta f\) is the observing bandwidth (MHz), \(W\) is the equivalent width of the pulse profile (ms), and \(P\) is the pulsar period (ms). 

\begin{figure} [!htp]
   \centering
   \includegraphics[width=9cm]{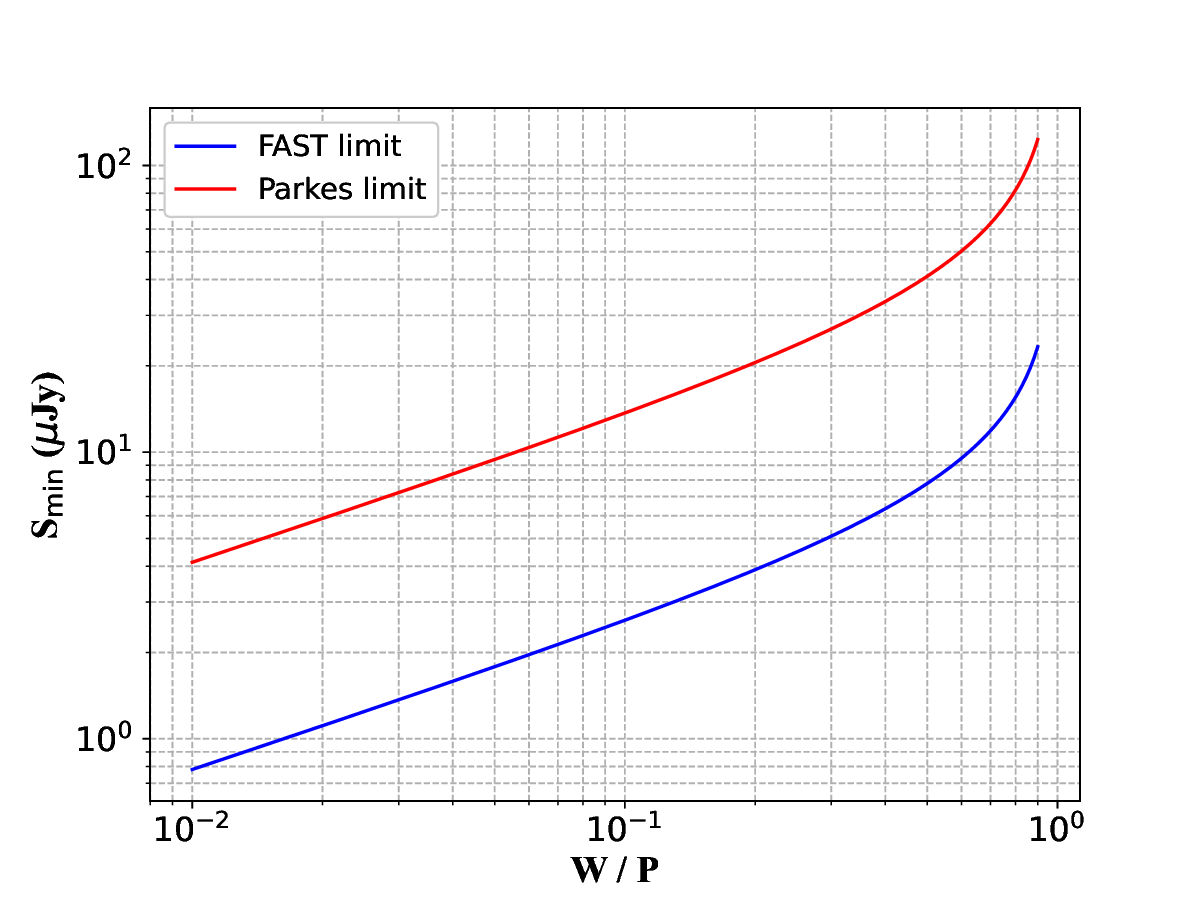}
   \caption{The theoretical upper limit of the radio flux density for the PULXs varying with the duty cycle. }
   \label{fig1} 
\end{figure}

For our observations with FAST, we used \(G = 16.7\) K Jy\(^{-1}\), \(T_{\text{sys}} = 25.7\) K, \(\Delta f = 400\) MHz, \(n_p = 2\), \(t_{\text{obs}} = 2400\) s, and \(\beta = 1\) \citep{jiang2020RAA}. For the Parkes telescope, the parameters used were \(G = 0.8\) K Jy\(^{-1}\), \(T_{\text{sys}} = 22\) K, \(\Delta f = 3328\) MHz, \(n_p = 2\), \(t_{\text{obs}} = 3300\) s, and \(\beta = 1.5\) \citep{Hobbs2020PASA}. We assumed the pulsed duty cycle (W/P) of range 0.01-0.9 and an S/N threshold of 7 for each case. The theoretical upper limit of the radio flux density are show in Figure \ref{fig1}.

\begin{figure} [!htp]
   \centering
   \includegraphics[width=9cm]{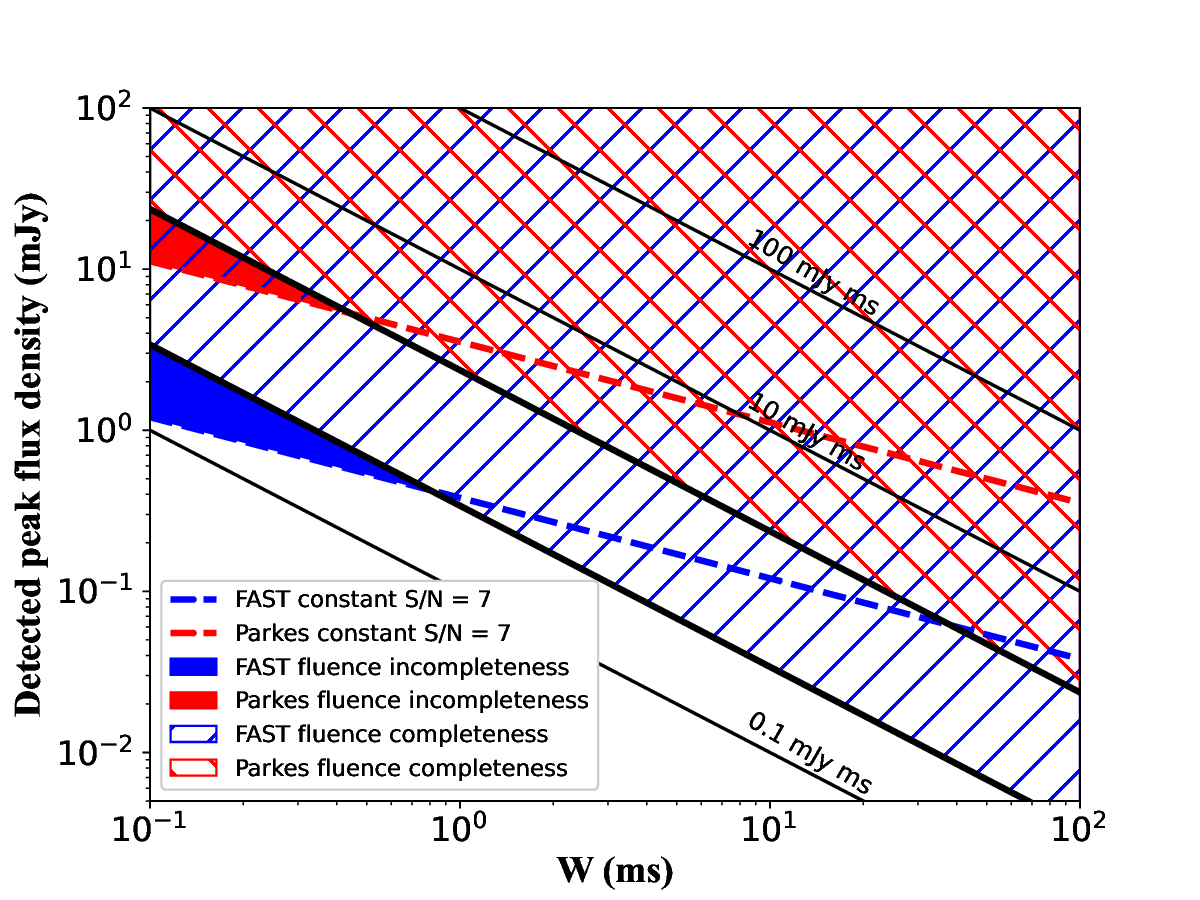}
   \caption{The flux density width parameter space. Lines of constant signal-to-noise (dashed) and constant  fluence (solid) are shown. In the diagonal area we have fluence completeness and in the shaded triangle we have fluence incompleteness.}
   \label{fig2} 
\end{figure}

The upper limit for a single pulse can be calculated using:
\begin{equation}
  S_{\text{SP}} = (S/N)_{\text{min}} \frac{\beta T_{\text{sys}}}{G} \sqrt{\frac{1}{n_p \Delta f W}}
  \label{eq2}
\end{equation}
Using the modified radiometer equation \ref{eq2} and the above S/N threshold, we calculated the upper limits for single-pulse flux density over pulse durations ranging from 0.1 to 100 ms. These calculated limits are illustrated in Figure \ref{fig2}.

\section{Discussion and Conclusions}\label{sec4} 

In this study, we conducted observations of eight PULXs using the FAST and Parkes telescopes, accumulating a total of 8 hours of observation for four PULXs using FAST and 11 hours for the other four PULXs using Parkes. We performed both single-pulse and periodic signal searches on the observational data, covering a DM range of 0–5000 pc cm$^{-3}$, aimed at detecting potential FRBs or pulsar-like emissions. 
Although no significant radio signals analogous to those of FRBs or pulsars were detected from these PULXs, our results place strict upper limits on the flux density of any potential radio emissions from these sources. 

PULXs with varying spin periods are expected to exhibit different duty cycles, with long-period sources likely having smaller duty cycles. To investigate how the duty cycle affects the upper limit of radio flux density (S$_{\mathrm{min}}$), we computed S$_{\mathrm{min}}$ across a range of duty cycles. Figure \ref{fig1} illustrates the theoretical variation of S$_{\mathrm{min}}$ as a function of the duty cycle for PULXs. Due to its higher sensitivity, the FAST achieves significantly lower detection limits compared to the Parkes telescope.
The Figure \ref{fig2} illustrates the parameter space of detected peak flux density versus observed pulse width, constrained by the single-pulse search limits for our observations of 8 PULXs. The dashed lines represent constant S/N = 7, while the solid lines indicate constant fluence. The regions shaded in red and blue denote the fluence completeness and incompleteness limits for the Parkes and FAST telescopes, respectively. In the fluence completeness region (diagonal area), the sensitivity is sufficient to detect single pulses if present, whereas in the fluence incompleteness region (shaded triangles), weaker or broader pulses might remain undetectable due to the telescopes' sensitivity thresholds.
Next, we discuss several possible reasons why FRBs or pulsar-like emissions have not been observed in PULXs.

For a pulsar in an accreting binary system, the radio pulsar will switch on when the accreting process ceases (see \citealt{1991PhR...203....1B} for a review), according to the recycling scenario, as evidenced by the discovery of transitional millisecond pulsars (MSPs), which switch between being a radio MSP and an accreting X-ray MSP~\citep{Archibald2009Sci, 2013Natur.501..517P, Bassa2014MNRAS, Papitto2022ASSL}. 
The spin period of the pulsar in PULXs ranges from 0.4\,s to 765.6\,s (see \citealt{Kaaret2017ARA&A} for a review), which is much larger than that of MSPs. However, since the pulsar in a PULX is in an accreting state, it shares some similar properties with accreting X-ray MSPs. 
Therefore, it is reasonable to assume that the periodic radio signals from neutron stars in PULXs are not detected due to the accretion process. 
Additionally, the companions of PULXs are thought to be massive stars~\citep{Rodr2020ApJ}, and the pulsed emission could be easily absorbed by the dense wind of these high-mass X-ray binaries, even if the radio pulsars were to switch on.

A short-lived relativistic outflows from accreting neutron stars could result in FRBs \citep{Sridhar2021ApJ}.
To generate the FRB, the outflows should be highly relativistic, and the jet should has extremely low luminosity and/or very high magnetization~\citep{Sridhar2021ApJ}. 
However, the environment for FRBs may be not clear for a PULX wind region. 
Even if the environment is clean and optically thin for FRB, the factors are strict to create observed FRBs.
PULXs are characterized by a narrow open funnel along the polar axis, defined by the disk's angular momentum, generating X-ray emissions~\citep{2006MNRAS.370..399B,Sridhar2021ApJ}.
The beaming factor for the X-ray emission of ULXs can be described as \citep{2009MNRAS.393L..41K}
\begin{equation}
\label{eq3}
f_{\rm b,X} =\left\{
\begin{array}{ll}
0.7, &  \dot{m}  \ll 10 \\
\frac{73}{\dot{m}^2}, &  \dot{m}  \gg 10 \\
\end{array} \right.
\end{equation}
where $\dot{m}$ is the Eddington ratio for accretion. 
In PULXs, the relativistic outflow responsible for powering the FRB will likely be confined to a similar solid angle, with a beaming factor larger than $f_{\rm b,X}$~\citep{Sridhar2021ApJ}. 
For the eight PULXs that we observed, the $f_{\rm b,X}$ are estimated as $\sim 1.8\times 10^{-5}$ to 0.3.
Such small beaming factor could account for the non-detection.

In terms of the null hypothesis that the progenitors of FRBs are periodically orbiting in binary system, our non-detection of both single pulses and radio pulsation from the PULXs can put some constraints on the periodicity and occurrence rates of FRBs associated with these systems. Observationally, the periodic activity found in FRB 20180916B (16.35 days; \citealt{Chime/FRB2020Natur.582}) and FRB 20121102A (156.9 days; \citealt{Rajwade2020MNRAS}) supports binary origins, as bursts could be tied to orbital phases. 
Moreover, the localization of FRB 20200120E to a globular cluster (GC) in M81 \citep{Kirsten+22Nat} further strengthens this scenario due to the prevalence of binary systems in such environments, though there has been only one GC-associated source identified until now. 
However, PULXs are high-mass X-ray binaries with massive stellar companions and are relatively young, whereas GC binaries are typically older systems with low-mass or degenerate companions.
PULXs, as common binary systems, provide a plausible probe to examine how common FRBs reside in binary systems and if long-term periodicity of those FRBs are caused by the orbital motions. In general, our results imply that FRB events are not ubiquitous in universal binary systems hosting magnetized neutron stars, likely requiring specific physical scenarios, even under super-Eddington accretion. Unless specific conditions, such as unique orbital configurations, magnetic field interactions, or episodic accretion events, one can hardly expect coherent emission processes required for FRBs' extremely high brightness temperature.

For ULX-like binaries, the conditions required for FRB production remain poorly understood, though it has been proposed that FRBs could originate from short-lived relativistic outflows \citep{Sridhar2021ApJ}.
\citet{2018Natur.562..233V} reported the first detection of a radio jet during the giant outburst of the PULX Swift J0243.6+6124, providing evidence that a strongly magnetized accreting neutron star can indeed launch a jet.
Subsequent observations revealed that these radio jets are transient, switching on and off within days during X-ray re-brightening episodes \citep{Eijnden2019MNRAS}.
The detection of radio jets suggests that the mechanisms governing radio emission in a super-Eddington accretion environment are highly complex.
High-cadence monitoring, particularly during X-ray outbursts or re-brightenings, could offer crucial insights into the interplay between accretion, jet formation, and radio emission in these systems.
We will keep monitoring these sources further using radio telescopes around the world, which may provide us more hints on the condition of coherent emission under super-Eddington accretion environment.

\section*{Acknowledgments}

This is work is supported by the National Natural Science Foundation of China (No. 12288102, No. 12203092, No. 12303042, No. 12041304, No. 12403058),  the Major Science and Technology Program of Xinjiang Uygur Autonomous Region (No. 2022A03013-3), the National SKA Program of China (No. 2020SKA0120100), the National Key Research and Development Program of China (No. 2022YFC2205202, No. 2021YFC2203502), the Natural Science Foundation of Xinjiang Uygur Autonomous Region (No. 2022D01B71), the Tianshan Talent Training Program for Young Elite Scientists (No. 2023TSYCQNTJ0024). 
This work made use of the data from the Five-hundred-meter Aperture Spherical radio Telescope, which is a Chinese national megascience facility, operated by National Astronomical Observatories, Chinese Academy of Sciences. The research is partly supported by the Operation, Maintenance and Upgrading Fund for Astronomical Telescopes and Facility Instruments, budgeted from the Ministry of Finance of China (MOF) and administrated by the Chinese Academy of Sciences (CAS).
This work made use of the data from FAST (Five-hundred-meter Aperture Spherical radio Telescope) (\url{https://cstr.cn/31116.02.FAST}). FAST is a Chinese national mega-science facility, operated by National Astronomical Observatories, Chinese Academy of Sciences. 
Murriyang, CSIRO’s Parkes radio telescope, is part of the Australia Telescope National Facility (\url{https://ror.org/05qajvd42}) which is funded by the Australian Government for operation as a National Facility managed by CSIRO. We acknowledge the Wiradjuri people as the Traditional Owners of the Observatory site.  We thank the referee for the valuable comments, which improved our manuscript.

\bibliography{references}{}

\bibliographystyle{aasjournal}

\end{document}